# Micro-sized cold atmospheric plasma source for brain and breast cancer treatment

Zhitong Chen[1], Li Lin[1], Qinmin Zheng[2], Jonathan H. Sherman[3], Jerome Canady[4], Barry Trink[5], Michael Keidar[1]*

[1]Department of Mechanical and Aerospace Engineering, The George Washington University, Washington, DC 20052, USA

[2]Department of Civil and Environmental Engineering, The George Washington University, Washington, DC 20052, USA

[3]Department of Neurosurgery, The George Washington University, Washington, DC 20052, USA

[4]Jerome Canady Research Institute for Advanced Biological and Technological Sciences, US Medical innovation LLC, Takoma Park, MD 20912, USA

[5]Department of Otolaryngology, School of Medicine, Johns Hopkins University, Baltimore, Maryland 21205, USA

---

* Corresponding Author:
E–mail address: zhitongchen@gwu.edu, keidar@gwu.edu

**Abstract**

Micro-sized cold atmospheric plasma (μCAP) has been developed to expand the applications of CAP in cancer therapy. In this paper, μCAP devices with different nozzle lengths were applied to investigate effects on both brain (glioblastoma U87) and breast (MDA-MB-231) cancer cells. Various diagnostic techniques were employed to evaluate the parameters of μCAP devices with different lengths such as potential distribution, electron density, and optical emission spectroscopy. The generation of short- and long-lived species (such as hydroxyl radical (•OH), superoxide ($O_2^-$), hydrogen peroxide ($H_2O_2$), nitrite ($NO_2^-$), et al) were studied. These data revealed that μCAP treatment with a 20 mm length tube has a stronger effect than that of the 60 mm tube due to the synergetic effects of reactive species and free radicals. Reactive species generated by μCAP enhanced tumor cell death in a dose-dependent fashion and was not specific with regards to tumor cell type.

**Key words:** Micro-sized, Cold atmospheric plasma, Reactive species, Breast cancer, Glioblastoma cancer, cancer therapy

**Introduction**

Cold atmospheric plasma (CAP) has been proposed as a novel therapeutic method for anticancer treatment, which can be applied to living tissues and cells[1,2]. CAP is a partially ionized gas that contain charge particles, reactive oxygen and nitrogen species (ROS and RNS), excited atoms, free radicals, UV photons, electric field, etc[3,4]. ROS and RNS, combined or independently, are well known to initiate different signaling pathways in cells and to promote oxidative stress[5,6]. Plasma-induced biological effects include damage lips, proteins, DNA, and induce apoptosis through plasma-generated ROS and RNS[7-10]. Moreover, many studies have reported both in vivo and vitro that plasma is a possible adjunct treatment in oncology as well as killing achieved for various types of cancers such as glioblastoma, breast cancer, bladder carcinoma, cervical carcinoma, skin carcinoma, pancreatic carcinoma, lung carcinoma, colon carcinoma, gastric carcinoma, melanoma and hepatocellular carcinoma[11-27].

In plasma medicine, jet plasma, corona discharge, and dielectric barrier discharge (DBD) have been used[28]. These types of plasma can be directly applied to skin cancers, while they are not applicable for more systemic cancer treatment. Some studies investigated the plasma device in the micro-sized to conduct the plasma species to the living animals[29]. However, their device just applied to xenografts tumors not systemic cancer treatment. Moreover, delivery of the plasma species is crucial to suppress tumor growth and assess efficiency of micro-sized plasma device. Hence, this study aims to design micro-sized cold atmospheric plasma devices with different lengths of nozzle in order to enhance delivery of reactive species and evaluate the efficiency of these devices on cancer therapy. Fig. 1 shows the potential applications of μCAP for brain and breast tumors in the future.

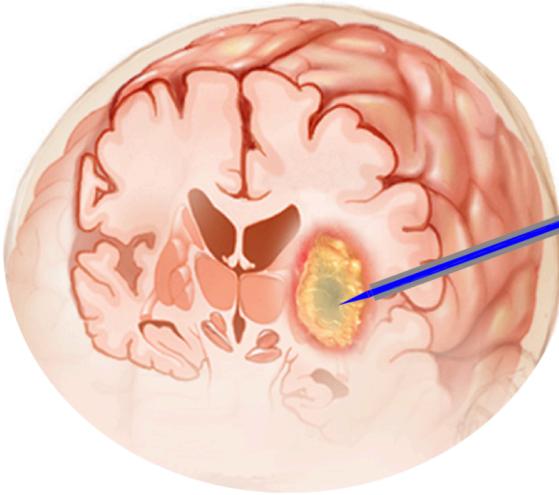 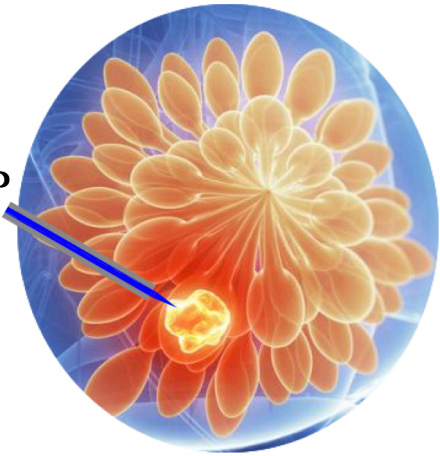

Figure 1. Potential applications of μCAP for brain and breast tumors

**Materials and Methods**

Fig. 2 depicts the schematic of the experiment setup including high voltage power (Fig. 2a) and μCAP devices (Fig. 2b). The high voltage power includes DC input, Trigger signal + MOSFET (switch), and the secondary output. In this work, the DC input was set at 5 V, square wave signal was obtained from the control unit (upper left in Fig. 2a), and a high voltage wave was obtained from the square wave signal through the transformer (upper right in Fig. 2a). The μCAP devices consist of a two-electrode (copper) assembly with a central powered electrode (1 mm in diameter) and a grounded outer electrode wrapped around the outside of a quartz tube (10 mm) as shown in Fig. 2b. The electrodes were connected to the secondary output of the high voltage transformer. The peak-peak voltage was approximately 8 kV and the frequency of the discharge was around 16 kHz (upper right in Fig. 2a). The secondary output of high voltage transformer was connected to the first input. At the end of a quartz tube, a $275 \pm 5$ μm inner diameter capillary tube (stainless steel) with 20 or 60 mm length was attached and insulated by epoxy. The feed gas for this study was industrial purity helium, which was injected into the quartz tube with a 0.2 L/min gas flow rate. Longer tube (60 mm) is needed to access deeper tumors in brain and breast. In this study, we are accessing effect of length to understand limitation of depth.

In this study, we are assessing the effect of tube length to understand limitation of depth. For instance, it is believed that a longer tube (60 mm) is needed to access deeper tumors in brain and breast. UV-visible-NIR, a range of wavelength 200-850 nm, was investigated on plasma to detect various RNS and ROS (nitrogen [$N_2$], nitric oxide [–NO], nitrogen cation [$N^{+2}$], atomic oxygen [O], and hydroxyl radicals [–OH]). The optical probe was placed at distance of 1.0 cm in front of the plasma jet nozzle. Data were then collected with an integration time of 100 ms.

A fluorimetric hydrogen peroxide assay Kit (Sigma-Aldrich) was used for measuring the amount of $H_2O_2$, according to the manufacturer's protocol. Briefly, 50 μl of standard curve, control, and experimental samples were added to 96-well flat-bottom black plates, and then 50 μl of Master Mix was added to each of well. The plates were incubated for 20 min at room temperature protected from light and fluorescence was measured by a Synergy H1 Hybrid Multi-Mode Microplate Reader at Ex/Em: 540/590 nm.

RNS level were determined by using a Griess Reagent System (Promega Corporation) according to the instructions provided by the manufacturer. Briefly, 50 μl of samples and 50 μl of the provided Sulfanilamide Solution were added to 96-well flat-bottom plates and incubated for 5-10 minutes at room temperature. Subsequently, 50 μl of the NED solution was added to each well and incubated at room temperature for 5-10 minutes. The absorbance was measured at 540 nm by Synergy H1 Hybrid Multi-Mode Microplate Reader.

XTT sodium salt ((2,3-bis(2-methoxy-4-nitro-5-sulfophenyl)-5-[(phenylamino)carbonyl]-inner salt-2H-tetrazolium, monosodium salt)) solution, purchased from Cayman chemical, was prepared by dissolving XTT power in DMEM. XTT sodium salt solution (100 μl per well, 500 μM) in a 96-well flat-bottom plate by μCAP for 5, 10, 30, 60, and 120 seconds. The gap between the outlet of μCAP and the surface of the samples was set at approximately 3 mm. As a control, untreated XTT sodium salt solution in triplicate were transferred to a 96-well flat-bottom plate. As a control, DMEM (100 μl per well) was treated with μCAP for 5, 10, 30, 60, and 120 seconds. The color change of XTT solution was used to indicate the presence of superoxide ($O_2^-$). A color change of XTT solution was measured by Hach DR 6000 uv vis spectrophotometer at 470 nm.

A MB solution was prepared by dissolving MB power in DMEM. MB solutions (100 μl per well, 0.01g/L) in a 96-well flat-bottom plate were treated by μCAP for 5, 10, 30, 60, and 120 seconds.

The gap between the outlet of μCAP and the surface of the samples was approximately 3 mm. As a control, untreated MB solutions in triplicate were transferred to a 96-well flat-bottom plate. The color change of methylene blue shows the presence of OH radicals via immediate and distinct bleaching of methylene blue dye (qualitatively analysis). The color change of the MB solution was measured as the absorbance at 664 nm by a Synergy H1 Hybrid Multi-Mode Microplate Reader. Human glioblastoma cancer cells (U87MG, Perkin Elmer) were cultured in Dulbecco's Modified Eagle Medium (DMEM, Life Technologies) supplemented with 10% (v/v) fetal bovine serum (Atlantic Biologicals) and 1% (v/v) penicillin and streptomycin (Life Technologies). Cultures were maintained at 37°C in a humidified incubator containing 5% (v/v) $CO_2$. The human breast cancer cell line (MDA-MB-231) was cultured in Dulbecco's Modified Eagle Medium (DMEM, Life Technologies) supplemented with 10% (v/v) foetal bovine serum (Atlantic Biologicals) and 1% (v/v) penicillin and streptomycin (Life Technologies). Cultures were maintained at 37 °C in a humidified incubator containing 5% (v/v) $CO_2$.

U87 and MDA-MB-231 cells were plated in 96-well flat-bottom microplates at a density of 3000 cells per well in 100 $\mu L$ of complete culture medium. Cells were incubated for 24 hours to ensure proper cell adherence and stability. On day 2, the cells were treated by He μCAP for 0, 5, 10, 30, 60, and 120 seconds. Cells were further incubated at 37°C for 24 and 48 hours. The cell viability of the glioblastoma and breast cancer cells were measured for each incubation time point with an MTT assay. 100 $\mu L$ of MTT solution (3-(4, 5-dimethylthiazol-2-yl)-2,5-diphenyltetrazolium bromide) (Sigma-Aldrich) was added to each well followed by a 3-hour incubation. The MTT solution was discarded and 100 $\mu L$ per well of MTT solvent (0.4% (v/v) HCl in anhydrous isopropanol) was added to the wells. The absorbance of the purple solution was recorded at 570 nm with a Synergy H1 Hybrid Multi-Mode Microplate Reader.

## Results and Discussion

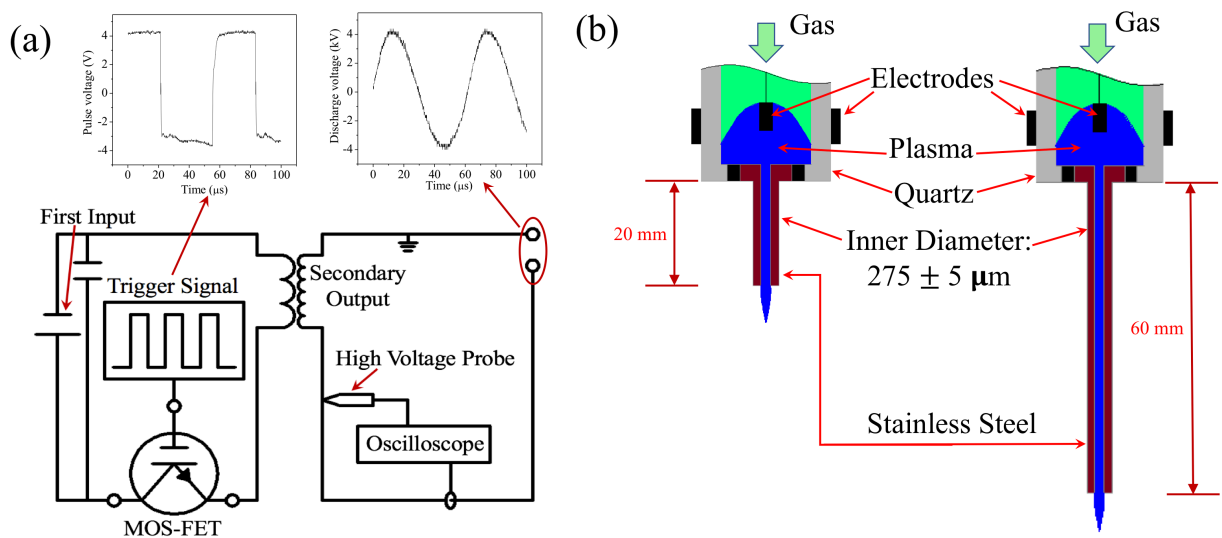

Figure 2. Schematic representation of the experiment setup including high voltage power part (a) and the micro-sized cold atmospheric plasma with 20 mm and 60 mm length of stainless steel tubes (b).

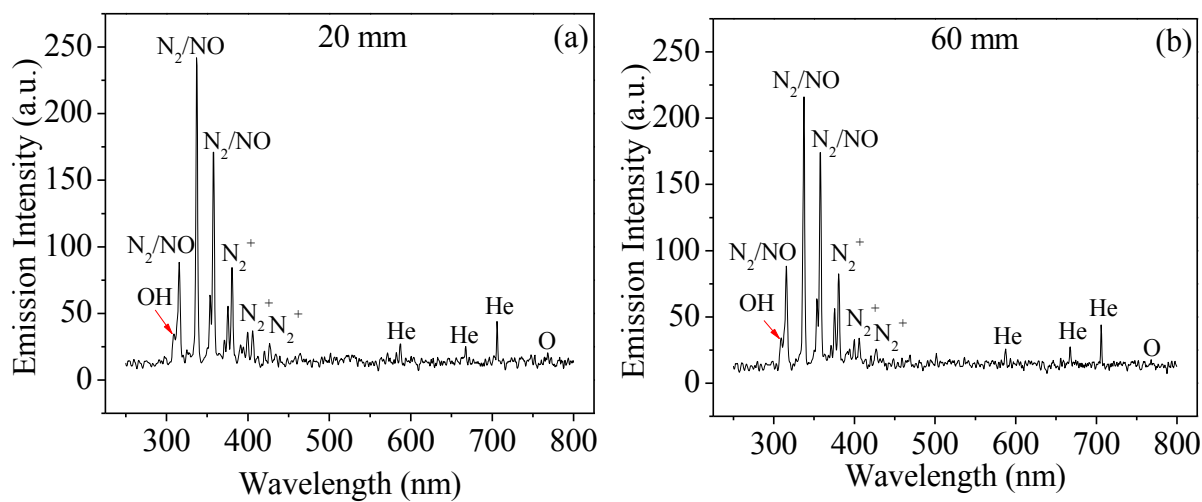

Figure 3. Optical emission spectrum detected from the He μCAP with 20mm (a) and 60 mm (b) length's tube using UV-visible-NIR, in the 250–850 nm wavelength range.

The reactive species generated by the *μ*CAP device with different micro-sized tube length are detected by optical emission spectroscopy, as shown in Fig. 3. The identification of the emission line and bands was performed mainly according to reference[30]. For 20 mm and 60 mm length devices, an $N_2$ second-positive system (315 nm, 337 nm 357 nm, and 380 nm) representing the photon emission intensity drops from the state $C^3\Pi u$ to $\beta^3\Pi g$ with different upper and lower vibration quantum numbers. There are very weak emission lines in the special range of 250-300 nm, which are detected as NO lines. The helium bands were assigned between 500 and 750 nm as shown in Fig. 3a and 3b. We also observed a high-intensity OH/O$_3$ peak at 309 nm for both 20 mm and 60 mm length devices. Atomic oxygen (O, including the ground state and all the excited states of atomic oxygen) was observed at 777 nm in both devices, which was believed to have a significant effect on cells and therefore a broad biomedical application. Micro-sized plasma is a complicated environment that combines the comprehensive effect of different ions and reactive species. The 60 mm µCAP has a bit less electron and species than 20 mm µCAP due to long distance delivery.

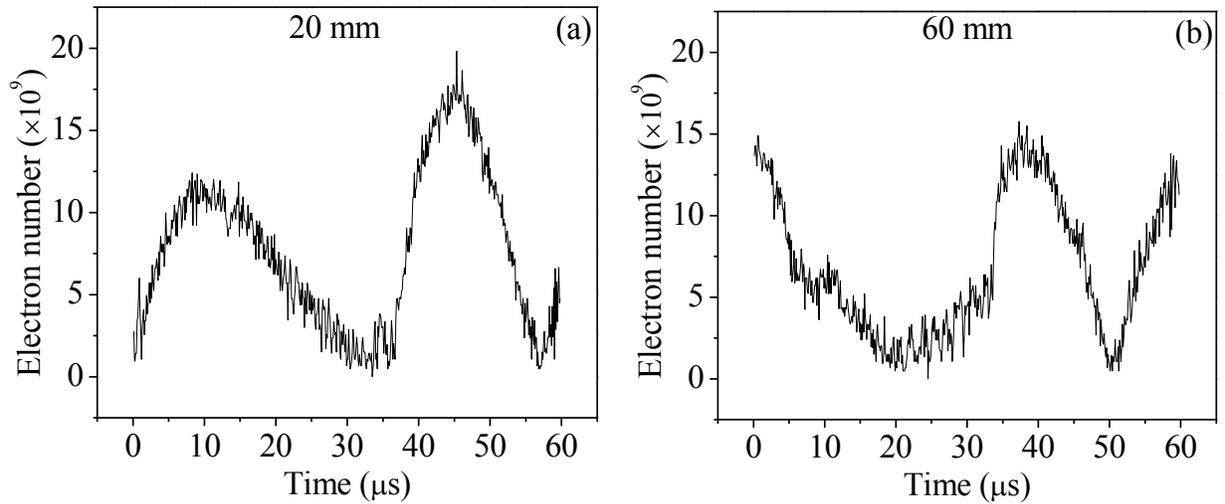

Figure 4 The electron number of 20 mm (a) and 60 mm (b) length µCAP

The experimental Rayleigh microwave scattering (RMS) system was described previosuly[19]. The detection of the scattered signal was accomplished using a homodyne scheme by means of an *I/Q* mixer, providing in-phase (*I*) and quadrature (*Q*) outputs. For the entire range of scattered signals, the amplifiers and mixer were operated in linear mode. The total amplitude of the scattered microwave signal was determined by: $U = \sqrt{I^2 + Q^2}$. We can calculate the total electron number in the plasma as $N_e = U(w^2 + v_m^2)/(2.82 \times 10^{-4} A v_m)$, where $w$ is the angular frequency, $v_m$ is the frequency of the electron-neutral collisions, and *A* is the proportionality coefficient[31]. The total electron number in the jet from μCAP with 20 mm and 60 mm is presented in Fig. 4a and 4b, and the total electron number for one discharge period is $4.60 \times 10^{12}$ and $4.04 \times 10^{12}$, respectively. A very small decrease of electron number has been detected in 60 mm μCAP comparing with 20 mm μCAP.

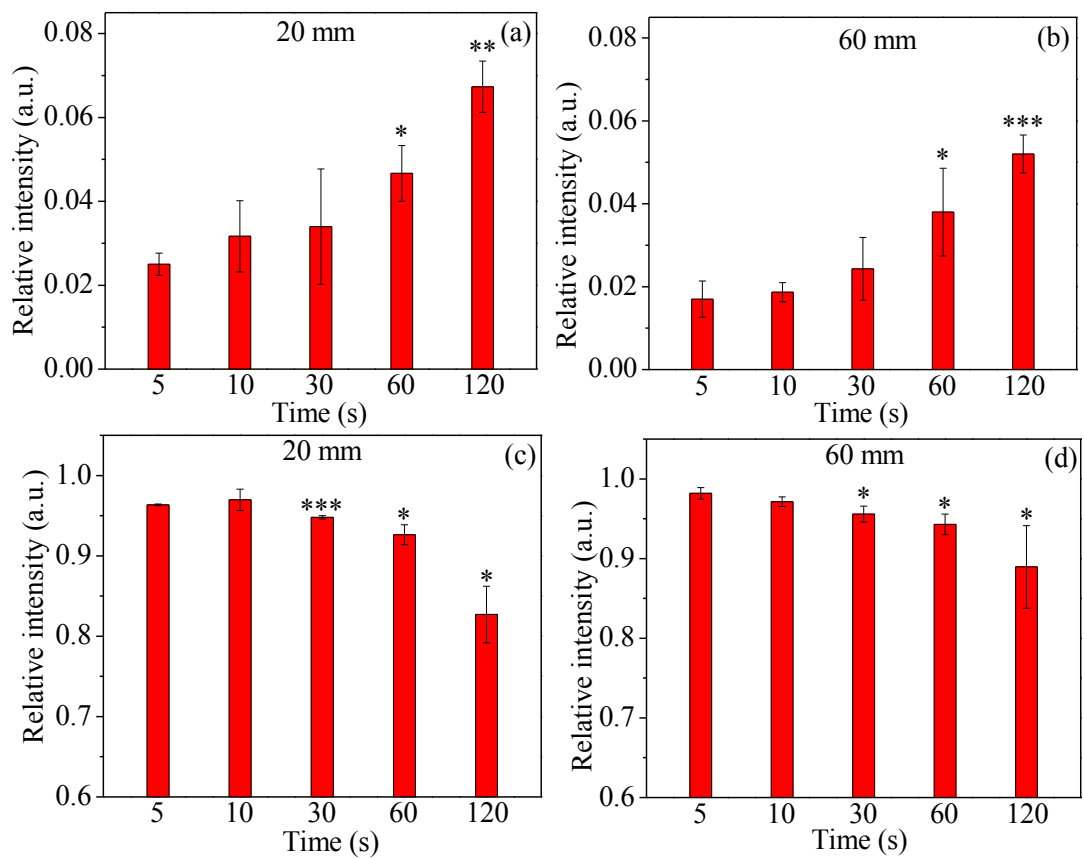

Figure 5. Relative $O_2^-$ and •OH concentration of 20 mm and 60 mm μCAP-treated DMEM. For relative $O_2^-$ concentration: (a) 20 mm and (b) 60 mm. For relative •OH concentration: (c) 20 mm and (d) 60 mm. Student t-test was performed, and the statistical significance compared to μ CAP 5 s treatment is indicated as * $p < 0.05$, ** $p < 0.01$, *** $p < 0.001$. (n = 3).

XTT solution was used to determine the relative concentration of superoxide ($O_2^-$). Superoxide radical reduced soluble formazans of the tetrazolium dye XTT[32,33]. Fig. 5a and 5b shows the relative superoxide concentration of 20 mm and 60mm μCAP treatment of DMEM. Relative intensity increases with treatment, which corresponds to the relative concentration of superoxide increasing with treatment. Comparing the 20 mm with 60 mm lengths, the 20 mm μCAP device produced a higher relative concentration of superoxide than the 60 mm device. Methylene blue (MB) was used to assess the relative concentration of hydroxyl radicals (•OH). It is known that MB reacts with •OH aqueous solutions, leading to a visible color change[34]. Fig. 5c and 5d shows that the relative MB concentration decreases with the treatment time of μCAP, suggesting that more •OH species are generated in DMEM (20 mm > 60 mm). Overall, these findings demonstrate that there is an increase in the relative concentration of $O_2^-$ and •OH as a function of μCAP treatment time.

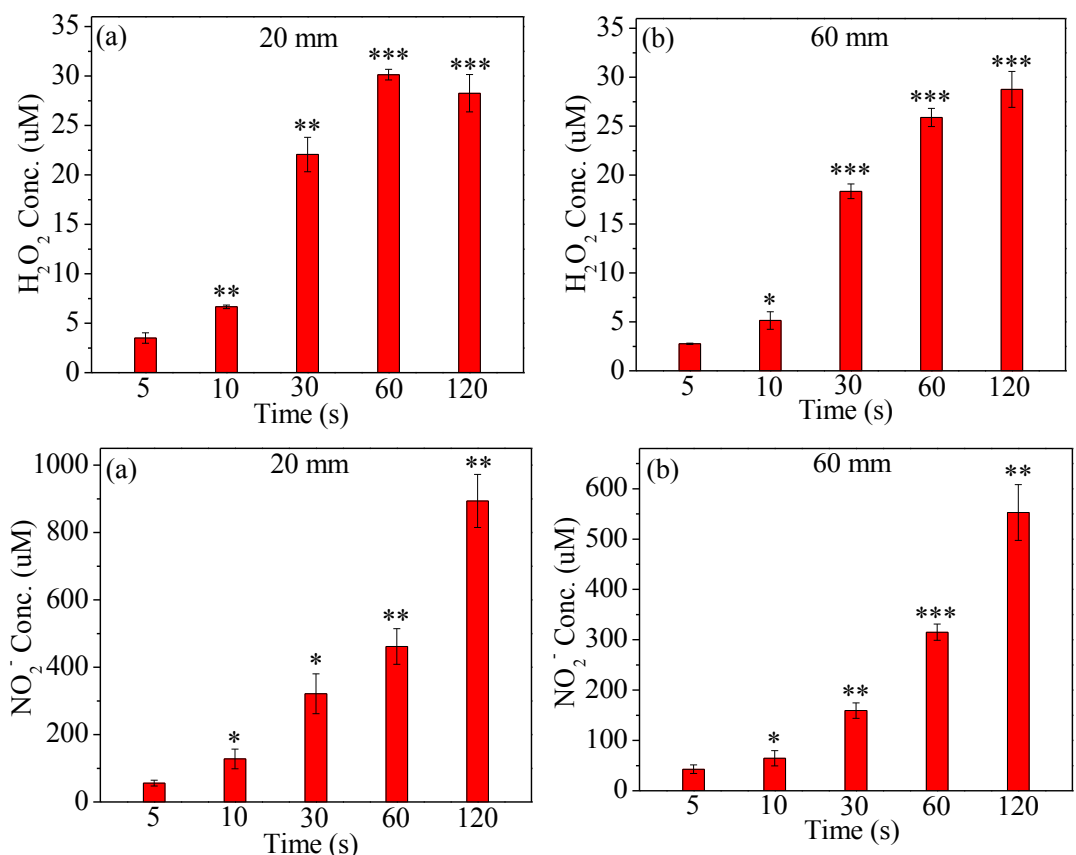

Figure 6. $H_2O_2$ and $NO_2^-$ concentration of 20 mm and 60 mm μCAP-treated DMEM. For $H_2O_2$ concentration: (a) 20 mm and (b) 60 mm. For $NO_2^-$ concentration: (c) 20 mm and (d) 60 mm. Student t-test was performed, and the statistical significance compared to μ CAP 5 s treatment is indicated as * $p < 0.05$, ** $p < 0.01$, *** $p < 0.001$. (n = 3).

DMEM treated by the 20 mm and 60 mm μCAP induced changes in the concentration of $H_2O_2$ and $NO_2^-$ as a function of the treatment time. These results are shown in Fig. 6 with concentrations produced by the 20 mm and 60 mm He μCAP devices. In Fig. 6a, the $H_2O_2$ concentrations produced by 20 mm He μCAP device increase with treatment time up to 60 seconds, but between 60 seconds and 120 seconds the concentration decreased. For the $H_2O_2$ concentration produced by 60 mm He μCAP increased with treatment time (In Fig. 6b). It means that the $H_2O_2$ concentration earlier reaches saturation in 20 mm length earlier than with the 60 mm length μCAP device. In Fig. 5, we know that He μCAP produces •OH and $O_2^-$ in DMEM, which are the two most important species in plasma-activated media. In particular, •OH reacting with •OH and $O_2^-$ reacting with $2H^+$ lead to $H_2O_2$ formation[35]. Both $NO_2^-$ concentrations of 20 mm and 60 mm increase with treatment time (in Fig. 6c and Fig. 6d), and $NO_2^-$ concentrations of 20 mm is much higher than 60mm. Comparing $NO_2^-$ concentration with the $H_2O_2$ concentration under same condition, $NO_2^-$ concentration is much higher than $H_2O_2$ concentration. A possible hypothesis for this result is that DMEM comprises over 30 components such as inorganic salts, amino acids and vitamins, and plasma might react with amino acids to form $NO_2^-$.

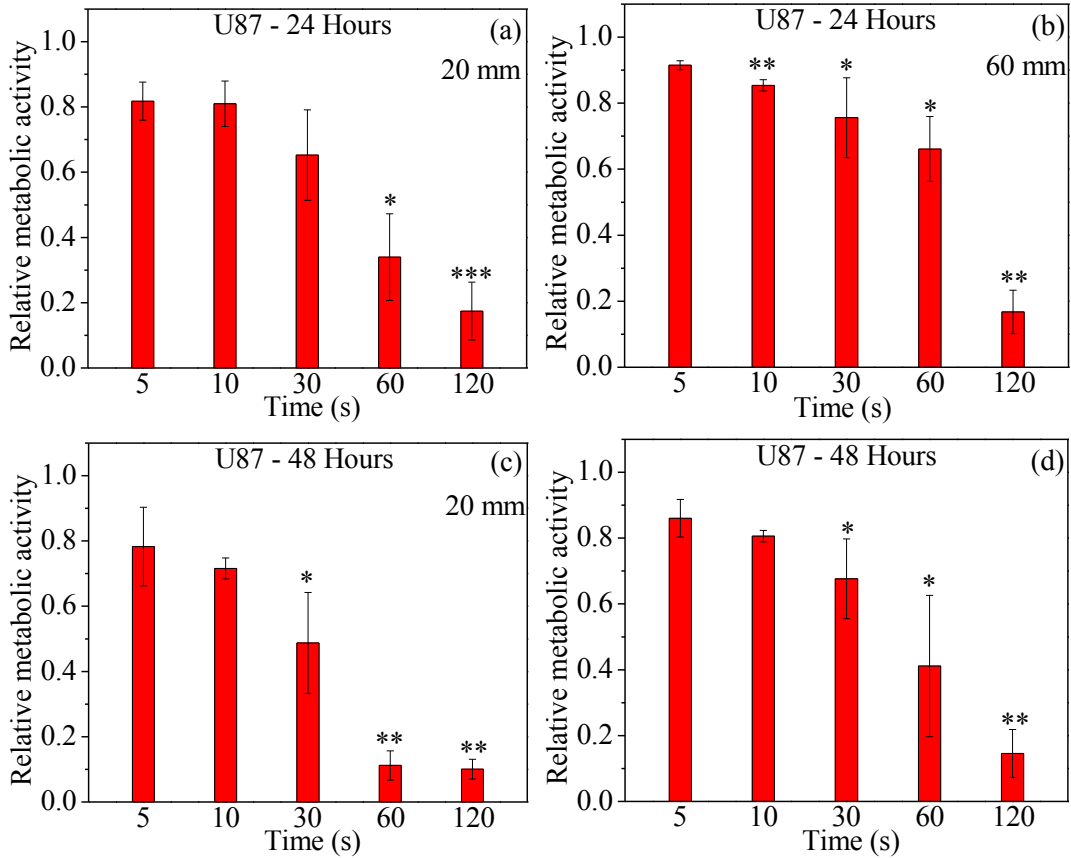

Figure 7. Cell viability of U87 after 24 and 48 hours' incubation with μCAP treatment with 20 mm and 60 mm length during 5, 10, 30, 60, and 120 seconds' treatment. Cell viability of U87 treated by 20 mm He μCAP at (a) 24-h incubation and (c) 48-h incubation. Cell viability of U87 treated by 60 mm He μCAP at (b) 24-h incubation and (d) 48-h incubation. The ratios of surviving cells for each cell line were normalized relative to controls (DMEM). Student t-test was performed, and the statistical significance compared to cells present in DMEM is indicated as $*p < 0.05$, $**p < 0.01$, $***p < 0.005$. (n=3)

Fig. 7 shows the cell viability of the brain (glioblastoma U87) cancer cells after 24 and 48 hours' incubation with μCAP during 5, 10, 30, 60, and 120 seconds' treatment with the 20 mm and 60 mm length μCAP device, respectively. For the 20 mm length μCAP treatment, the cell viability of brain cancer cells was lower than that of the 60 mm length at each treatment duration (from 5 to 60 seconds), and dropped with increasing treatment time. For both 20 mm and 60 mm, 120 seconds' treatment has similar effect on cell viability of U87 cancer cells. For 48 hours' incubation under 20 mm μCAP treatment, 60 and 120 seconds' duration has similar effect on cell viability. Thus, overall conclusion is that 60mm tube can still produce reactive species while allowing access to deeper tumors.

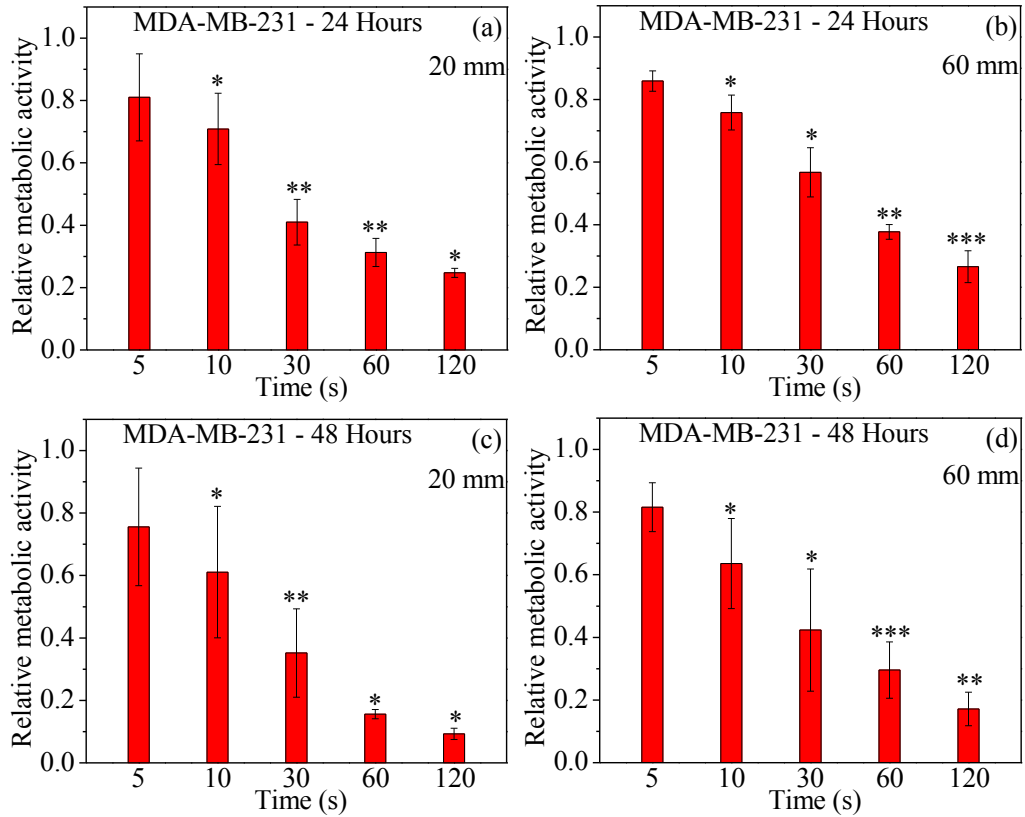

Figure 8. Cell viability of MDA-MB-231 after 24 and 48 hours' incubation with μCAP treatment with 20 mm and 60 mm length during 5, 10, 30, 60, and 120 seconds' treatment. Cell viability of MDA-MB-231 treated by 20 mm He μCAP at (a) 24-h incubation and (c) 48-h incubation. Cell viability of MDA-MB-231 treated by 60 mm He μCAP at (b) 24-h incubation and (d) 48-h incubation. The ratios of surviving cells for each cell line were calculated relative to controls (DMEM). Student t-test was performed, and the statistical significance compared to cells present in DMEM is indicated as *$p < 0.05$, **$p < 0.01$, ***$p < 0.005$. (n=3)

Fig. 8 shows the cell viability of the breast (MDA-MB-231) cancer cells after 24 and 48 hours' incubation with μCAP treatment with the 20 mm and 60 mm length μCAP devices during 5, 10, 30, 60, and 120 seconds' duration. For both 20 mm and 60 mm μCAP treatment, the cell viability after 24 and 48 hours' incubation dropped with increasing treatment time. For 20 mm μCAP treatment, the cell viability of breast cancer cells was lower than that of the 60 mm length at each treatment duration.

The direct plasma jet irradiation is limited to the skin and it can also be invoked as a supplement therapy during surgery as it only causes cell death in the upper three to five cell layers. However, the current cannulas from which the plasma emanates are too large for intracranial applications. Thus, we developed a micro-sized plasma devices with 20 mm and 60 mm length stainless steel tubes, which both can achieve effective killing of brain and breast cancer cells. This preliminary study offers significant potential for new treatment applications. Numerous studies reported plasma-induced apoptosis in cancer cells due to plasma-generated various reactive species[1,36,37]. Plasma generates various kinds of ROS and RNS, including hydrogen peroxide ($H_2O_2$), ozone ($O_3$), hydroxyl radical (•OH), atomic oxygen (O), superoxide ($O_2^-$), nitric oxide (NO) and peroxynitrite anion ($ONOO^-$), singlet delta oxygen ($O2(^1\Delta g)$), nitrite ($NO_2^-$)[37,38] and are displayed in Fig. 3. In this paper, we have specifically measured relative concentrations of $O_2^-$ and •OH (short-lived species, Fig. 5) and the concentration of $H_2O_2$ and $NO_2^-$ (long-lived species, Fig. 6). The relative concentration of $O_2^-$ treated by μCAP with 20 mm and 60 mm increases with treatment time (Fig. 5a and 5b). $O_2^-$ can activate mitochondrial-mediated apoptosis by changing mitochondrial membrane potential and simultaneously up-regulates pro-apoptotic genes and down-regulates anti-apoptotic genes for activation of caspases resulting in cell death[39]. Fig. 5c and 5d shows the relative concentration of •OH in DMEM treated by μCAP with 20 mm and 60 mm also increases with treatment time. •OH derived amino acid peroxides can contribute to cell injury because •OH itself and protein (amino acid) peroxides are able to react with DNA, thereby inducing various forms of damage[40]. Compared with cell viability of both cancer lines, the trend of cell death can be partly attributed to the increase of $O_2^-$ and •OH concentration with treatment time. On the other hand, the 20 mm μCAP device shows higher relative concentrations of $O_2^-$ and •OH, such that the 20 mm μCAP device is more effective in killing both cancer cell lines than the 60 mm μCAP device. Fig. 6 shows $H_2O_2$ and $NO_2^-$ concentration of the 20 mm and 60 mm μCAP-treated DMEM. $H_2O_2$ can induce cell death by apoptosis and necrosis, while $NO_2^-$ are known to induce cell death via DNA damage[36,41]. Thus, the synergism of $H_2O_2$ and $NO_2^-$ might be an important factor in cancer cells killing efficiency.

Several methods are now being used for the cancer treatment such as chemotherapy, surgery, and radiotherapy[42-44]. The conventional methods have some disadvantages such as low rapidity, high cost, and adverse effects. However, plasma treatment may overcome these disadvantages of the traditional treatments. Currently, plasma can be directly applied to skin cancers, while it is not applicable for more systemic cancer treatment. However, we developed novel μCAP with 20 mm and 60 mm length can be considered as a local treatment tool and does not exert the systemic therapeutic effects like chemical drugs, meanwhile removing limits of plasma itself. Overall, the above results and discussion indicate that both μCAP with 20 mm and 60 mm length might be useful and should be considered in a clinical medical application.

**Conclusions**

In this work, we showed that the newly developed micro-sized cold atmospheric plasma (μCAP) device with 20 mm and 60 mm length stainless steel tubes induce the production of reactive species and radicals in culture medium. There is an increase in the concentration of $O_2^-$, •OH, $H_2O_2$, and $NO_2^-$ as a function of μCAP treatment time, which matches the trend of cell viability of two cancer cells with μCAP treatment time. A synergistic effect of short- and long-lived species present in the plasma treating DMEM is suspected to play a key role in cell death. Both the 20 and 60 mm length devices have s significant effect on both U87 and MDA-MB-231 cancer cell viability, allowing access to both superficial and deeper tumors. The results of this study suggest a possibility for clinical applications of this micro-μCAP device on brain and breast tumors. Future work looks to utilized the micro-μCAP device inside the patient's body.